\documentclass[11pt]{article}
\usepackage{latexsym}
\usepackage{amssymb}

\newcommand{\C}{\mathbb{C}}
\newcommand{\CP}{\mathbb{CP}}

\newcommand{\R}{\mathbb{R}}

\renewcommand{\d}{\mathrm{d}}

\def\be{\begin{equation}}
\def\ee{\end{equation}}

\def\Sm{\Sigma}

\def\Om{\Omega}
\def\Th{\Theta}
\def\om{\omega}

\def\p{\partial}

\def\a{\alpha}

\def\O{{\cal O}}

\newtheorem{theo}{Theorem} 
\newtheorem{defi}{Definition}
\nonumber
\begin{document}
\title{Null--K\"ahler structures, Symmetries and Integrability }
\author{Maciej Dunajski\\
Department of Applied Mathematics and Theoretical Physics, \\
Cambridge University,\\
Wilberforce Road, Cambridge, CB3 0WA, UK
\and
Maciej Przanowski\\ 
Institute of Physics,\\
Technical University of Lodz,\\
Wolczanska 219, 90--005 Lodz, Poland\\and\\ 
Departamento de Fisica, CINVESTAV,\\
Mexico D.F., Mexico.
}
\date{} 
\maketitle
We feel honoured to be able to make this small 
contribution to the celebration of  Jerzy Pleba\'nski's 75th birthday.
Pleba\'nski has presumably regarded his work on complex relativity as a 
step towards producing general solutions to the Einstein equations
on a real Lorentzian manifold.
No one in the mid-seventies could expect that his contributions 
to the field would underlie the relation  between  twistor descriptions of 
anti-self-dual conformal structures and 
integrable models!

The main focus of this paper will be a $(2, 2)$ signature metric in 
Pleba\'nski's form
\be
\label{nkmetric}
g=\d w\d x+\d z \d y-\Th_{xx}\d z^2-\Th_{yy}\d w^2+2\Th_{xy}\d w\d z.
\ee
Here $(w, z, x, y)$ are local coordinates in an open ball in $\R^4$, and
$\Th:\R^4\longrightarrow \R$ is an arbitrary real analytic function.
Not all $(2, 2)$ inner--products can be put in this form even locally.
To understand the local constraint imposed on $g$ by (\ref{nkmetric})
let us make the following
\begin{defi}
A null-K\"ahler structure on a real four-manifold ${\cal M}$ 
consists of an inner
product $g$ of signature $(++--)$ and a real rank-two endomorphism $
N:T{\cal M}\rightarrow T{\cal M}$ parallel 
with respect to this inner product such that
\[
N^2=0,\qquad\mbox{and}\qquad g(NX, Y)+g(X, NY)=0
\]
for all $X, Y \in T{\cal M}$.
\end{defi}
Consider the splitting  $T_{\C}{\cal M}\cong S_+\otimes S_-$, 
where $S_+$ and $S_-$ are complex two-dimensional spin bundles. 
The isomorphism ${\Lambda^2}_+({\cal M})\cong {\mbox{Sym}}^2(S_+)$ between 
the bundle of self-dual two-forms and the symmetric tensor product of 
two spin bundles implies that the  existence of a null--K\"ahler structure is
in four dimensions equivalent to the existence of a parallel real spinor.
The Bianchi identity implies the vanishing of the curvature scalar.
Null--K\"ahler structures are special cases of conformally recurrent 
structures investigated in \cite{PP89}.
In \cite{B00} and \cite{D02} it was shown that null--K\"ahler structures are
locally given by one arbitrary  function of four variables, and 
admit a canonical form (\ref{nkmetric}) with 
$N=\d w\otimes\p/\p y-\d z\otimes\p/\p x$.  

Further conditions can be imposed on the curvature of $g$ 
to obtain non--linear PDEs for the 
potential function $\Theta$
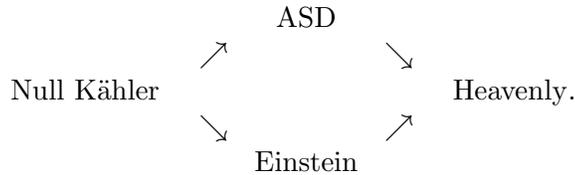
\[
\mbox{Null K\"ahler}\begin{array}{rcccl}
&& \mbox{ASD}  &&\\
&\nearrow&&\searrow &\\
&&&&\\
&\searrow&&\nearrow &\\
&&\mbox{Einstein}  &&
\end{array}
\mbox{Heavenly}.
\]
Define 
\be
\label{nk1}
f:=\Th_{wx}+\Th_{zy}+\Th_{xx}\Th_{yy}-\Th_{xy}^2
\ee
\begin{itemize}
\item The Einstein condition implies that
\be
\label{HHspace}
f=xP(w, z)+yQ(w, z)+R(w, z),
\ee
where $P, Q$ and $R$ are arbitrary functions of $(w, z)$. 
In fact the number of the arbitrary functions 
can be reduced down to one by redefinition of
$\Th$ and the coordinates, but for reasons which 
will become clear in the last section 
we prefer the form (\ref{HHspace}).
This is the hyper--heavenly equation
of Pleba\'nski and Robinson \cite{PR76} for non--expanding metrics of type
$[N]\times$[Any]. (Recall that $({\cal M},  g)$ is called hyper--heavenly if
the self--dual Weyl spinor is algebraically special).
The solvability of equation (\ref{HHspace}) will be discussed in the 
last section of this paper.
\item The conformal anti--self--duality (ASD) condition implies
a 4th order PDE for $\Theta$
\be
\label{md}
\square f=0,
\ee
where $\square$ is the Laplace--Beltrami operator defined by the metric $g$.
This equation is integrable: 
It admits a Lax pair and its solutions can in principle be found 
by twistor methods \cite{D02}.
\item Imposing both conformal ASD and Einstein condition implies (possibly
after a redefinition of $\Theta$) that $f=0$, which yields the
celebrated second heavenly equation of Pleba\'nski \cite{Pl75}
\be
\label{secondh}
\Th_{wx}+\Th_{zy}+\Th_{xx}\Th_{yy}-\Th_{xy}^2=0.
\ee
\end{itemize}
Many lower--dimensional integrable systems can be 
obtained form (\ref{secondh}) or (\ref{md}) if the associated metrics admit
symmetries. The analysis of  symmetry  
reductions can be made coordinate independent 
and more systematic by introducing some geometry on the space of orbits of the
Killing vector. The relevant structure consists of a conformal structure 
compatible with a torsion free connection. The constraints induced by 
(\ref{secondh}) or (\ref{md}) imply the Einstein--Weyl (EW) 
equations.
The study of these equation goes back to Cartan \cite{C43}. 
In the next section they will be presented in  a modern language 
of Hitchin \cite{Hi82}. We shall then  
review  the special classes of EW spaces, and their relation to 
solutions of the heavenly equation (\ref{secondh}). 
In the last section we shall
address the question of integrability of the non--ASD equation
(\ref{HHspace}). It will be shown that Null K\"ahler Einstein metrics
with symmetry preserving the Null-K\"ahler form locally depend on
solutions to the variable coefficient 
dispersionless Kadomtsev--Petviashvili equation. 
\section{Einstein--Weyl geometry and symmetry reductions}
Let $M$ be an $n$-dimensional manifold with a torsion-free connection $D$,
and a conformal structure $[h]$ which is compatible with $D$ in a sense
that
\[
Dh=\om\otimes h
\]
for some one-form $\om$.
Here $h\in[h]$ is a representative metric in a conformal class. If 
we  change this representative by $h\rightarrow \psi^2 h$, 
then $\om\rightarrow \om +2\d \ln{\psi}$, where $\psi$ is a 
non-vanishing function on $M$. 
The space of oriented $D$--geodesics in $M$ is a manifold $\cal Z$ 
of dimension $2n-2$. It can be identified with a quotient space
of the projectivised tangent bundle $P(TM)$ by the geodesic spray.
There exists a fixed point free 
map $\tau:{\cal Z}\longrightarrow {\cal Z}$ which reverses an
orientation of each geodesics.

To describe a tangent space to $\cal Z$ at
the geodesic $\gamma(t)$ take a curve of geodesics $\gamma(s, t)$ with
$\gamma(0, t)=\gamma(t)$ and consider the Jacobi vector field
\[
V=\frac{\p\gamma}{\p s}|_{s=0}.
\]
The $(2n-2)$ dimensional tangent space $T_{\gamma}{\cal Z}$ is then just the space of solutions
to the Jacobi's equation
\be
\label{jacobi}
(D_U)^2V+R(V, U)U=0
\ee
modulo vector fields tangent to $\gamma$. Here $U=\d \gamma/\d t$, and
$R$ is the curvature tensor of $D$ defined by
\[
R(U, V)W=[D_U, D_V]W-D_{[U, V]}W,\qquad U, V, W\in TM.
\]
Note that in general the Ricci tensor constructed out of 
$R$ is not symmetric, and its skew part is proportional
to $\d \om$. 

Now consider the special case of three-dimensional Weyl manifolds, and define
the almost-complex structure on ${\cal Z}$ by
\[
J(V)=\frac{U\times V}{\sqrt{h(U,U)}},
\]
where $\times$ is the usual vector product on $\R^3$.

If $V$ is a Jacobi field orthogonal to $U$ then $J^2=-$Id. 
The solution space to (\ref{jacobi}) is $J$--invariant if 
$([h], D)$ satisfy the conformally invariant Einstein--Weyl equations
\be
\label{EW}
R_{(ab)}=\frac{1}{3}rh_{ab},\qquad a, b, ... =1, 2, 3.
\ee
Here $R_{(ab)}$ is the symmetrised Ricci tensor of $D$, and $r$ is the
Ricci scalar. If equations (\ref{EW}) are satisfied, then $J$ is automatically
integrable. In fact we have the following result
\begin{theo}[Hitchin \cite{Hi82}]
\label{thhitchin}
There is a one-to-one correspondence between local solutions 
to the Einstein--Weyl equations {\em(\ref{EW})}, and complex surfaces 
(twistor spaces) equipped
with a fixed-point free anti-holomorphic involution $\tau$, and a 
$\tau$-invariant rational curve with a normal bundle $\O(2)$.
\end{theo}
The EW space can be completely reconstructed form the twistor data.
Since $H^0(\CP^1, \O(2))=\C^3$, and $H^1(\CP^1, \O(2))=0$
we can use Kodaira's theorem. The EW space is a space of those
$\O(2)$ curves which are $\tau$-invariant. The family of such curves
passing through  a given point (and its conjugate) is a geodesic of a 
Weyl connection  of $D$. 
To construct a conformal structure $[h]$ consider a point on a $\tau$-
invariant $\O(2)$ curve $L_p$. This point represents a point in a sphere
of directions $(T_pM-0)/\R^+$, and the conformal structure on $L_p$ 
induces a quadratic conformal structure in $M$.
\subsection{Special shear--free geodesic congruences}

Recall that a geodesic congruence $\Gamma$ in a region
in $\hat{M}\subset M$ is a set of geodesic, one through each point
of ${\hat M}$. Let $V$ be a generator of $\Gamma$ (a vector field tangent
to $\Gamma$). The geodesic condition $V^aD_aV^b\sim V^b$ implies 
$D_aV^b={M_a}^b+A_aV^b$ for some $A_a$, where ${M_a}^b$ is orthogonal to 
$V^a$ on both indices. Consider the decomposition of $M_{ab}$
\[
M_{ab}=\Om_{ab}+\Sm_{ab}+\frac{1}{2}\theta \hat{h}_{ab}
\]
The shear $\Sm_{ab}$  is trace-free and symmetric.
The twist $\Om_{ab}$ is anti-symmetric, and the divergence $\theta$ is 
is a weighted scalar..
Here $\hat{h}_{ab}=||V||^2h_{ab}-V_aV_b$ is an orthogonal projection of $h_{ab}$.
The 
shear-free geodesics congruences (SFC) exist on any Einstein--Weyl space.
This follows from a three-dimensional version of Kerr's theorem which 
states that SFCs correspond to  
to holomorphic curves in ${\cal Z}$.
On the other hand imposing conditions on twist and divergence of a congruence 
gives restrictions on EW structures, 
and can be used to reduce the EW equations to some known and
new integrable equations. This method was first applied in \cite{T95}.
The general theory of SFC and its relation to EW geometry was developed in
\cite{CP}.
\begin{itemize}
\item
Vanishing of the twist of an SFC implies 
existence of a foliation of an EW space by surfaces orthogonal 
to the congruence.  
It follows from the shear-free condition that these 
surfaces are equipped with a conformal structure. The EW structure can be
locally put in the form
\[
h=e^U(\d x^2+\d y^2)+\d t^2,\qquad \om=2U_t\d t,
\]
Here $(x, y)$ are isothermal coordinates on the surfaces, $\p/\p t$ 
is normal to the surfaces, and $U=U(x, y, t)$ is a function. 
The EW equations reduce \cite{T95} 
to the Boyer--Finley--Pleba\'nski (BFP) equation
\be
\label{BFP}
U_{xx}+U_{yy}+{(e^U)}_{tt}=0.
\ee
The preferred congruence is given by $\d t$ in the above coordinates.
The system of geodesics $(x, y)=$const equipped with two possible 
orientations becomes a pair of complex curves ${\cal D}$ and
$\tau({\cal D})$ in ${\cal Z}$. LeBrun \cite{L91} shows that the divisor class
${\cal D}+\tau({\cal D})$ represents the line bundle $\kappa^{-1/2}$,
where $\kappa\longrightarrow {\cal Z}$ is the canonical line bundle 
(the bundle of holomorphic two-forms). 
\item
The existence of a parallel congruence implies \cite{DMT00} the
existence of a local coordinate system such that
\[
h=\d y^2-4\d x\d t-4U\d t^2,\qquad \om=-4U_x\d t,
\]
and the EW condition reduces to the dispersion-less
Kadomtsev--Petviashvili (dKP) equation 
\be
\label{dkp}
(U_t-UU_x)_x=U_{yy}.
\ee
If $U(x, y, t)$ is a smooth real function of real variables then
the conformal structure 
has signature $(++-)$. The real structure $\tau$ on ${\cal Z}$
differs form the one considered in  Theorem \ref{thhitchin}. 
Now $\tau$ fixes an equator on each $\CP^1$ and interchanges upper and lower
hemisphere.
 One can verify that the  vector $\p/\p x$ 
is a real null vector, covariantly constant in the Weyl connection, and  
with weight $-1/2$. Covariantly constant real null vector givers rise
to a parallel real weighted spinor, and finally to a preferred
section of $\kappa^{-1/4}$ in ${\cal Z}$.
\item The existence of the divergence-free SFC implies \cite{D03} that
locally the EW structure is given by
\[
h=(\d y+U\d t)^2-4(\d x+W\d t)\d t,\qquad 
\om =U_x\d y+(UU_x+2U_y)\d t,
\]
where $U(x, y, t)$ and $W(x, y, t)$ satisfy a system  of quasi-linear PDEs 
\be
\label{PMA}
U_t+W_y+UW_x-WU_x=0,\qquad U_y+W_x=0.
\ee
The the preferred congruence $\d t$ is shear free, 
and its  divergence $D*(\d t)$ vanishes. The corresponding
twistor space ${\cal Z}$ fibres holomorphically over $\CP^1$ \cite{CP}.
\end{itemize}
It is interesting to note that equations (\ref{BFP}, \ref{dkp}, 
\ref{PMA}) are integrable in more than one sense 
as they possess infinitely many
hydrodynamic reductions \cite{FK03,GMM}.
\subsection{Examples of solutions}
Equations (\ref{BFP}) and (\ref{dkp}) are equivalent to 
\be
\label{gdkp}
\d\ast_h \d U=0,
\ee 
where $\ast_h$ is the Hodge operator taken with respect to the corresponding
EW conformal structure (equation (\ref{PMA}) also implies (\ref{gdkp}),
but the converse does not hold in general). 

Equations which can be written in the form (\ref{gdkp}) may be reduced to ODEs
by a `central quadric' ansatz.
The ansatz is to seek solutions constant on central quadrics 
or equivalently to seek a matrix $M_{ab}(U)$ so that
a solution of (\ref{gdkp}) is determined implicitly  by
\be
\label{quadricA}
M_{ab}(U)x^ax^b=C,
\ee
where $x^a=(x, y, t)$, and $C=$const.
The general method of reducing this condition to an ODE is described in 
\cite{DT02}. Although the ansatz leads to ODEs, the resulting solutions to
(\ref{gdkp}) are not group invariant.

 In the case of the BFP equation  
this ODE reduces to Painlev\'e III \cite{T95}, and
in the case of dKP the ODE reduces to Painlev\'e I or II \cite{DT02}. 
The details of the dKP case are as follows:
\begin{itemize}
\item
If $({M}^{-1})_{33}\neq 0$ then (\ref{quadricA}) becomes
\begin{eqnarray*}
&&{x}^{2}v-
{y}^{2}w\left (wv-(\a-1/2)\right )+\frac{1}{2}\,{t}^{2}\left (\left (\a-1/2
\right )^{2}+4\,wv\left (wv-(\a-1/2)\right )+2\,{v}^{3}\right )
\nonumber\\
&&+xy\left (\a-1/2\right )-ytv\left (\a-1/2\right )-2\,tx{v}^{2}=
C(2wv-(\alpha-1/2))^2,
\end{eqnarray*}
where $\a$ is a constant parameter,
\[
v=\frac{1}{2}\dot{w}(U)-w(U)^{2}-U,
\]
and $w(U)$ satisfies Painlev\'e II
\[
\label{rescalPII}
\frac{1}{4}\ddot{w}=2w^3+2wU+\a.
\]
\item
If $({M}^{-1})_{33}=0$ and $({M}^{-1})_{23}\neq 0$  then
(\ref{quadricA}) becomes
\[
x^2+w^2y^2
-w\Big(\frac{{\dot w}^2}{4}-4w^3\Big)t^2-4xtw^2
+2wxy+\Big(\frac{{\dot w}^2}{4}-4w^3\Big)yt=C\dot{w}^2,
\]
where $w(U)$ satisfies Painlev\'e I
\[
\ddot{w}/4=6w^2+2U.
\]
\item
Finally if $({ M}^{-1})_{33}=({M}^{-1})_{23}= 0$  then
dKP reduces to a  linear equation.
\end{itemize}
\subsection{Heavenly spaces with symmetry}
A link between three-dimensional EW geometry and symmetries of the 
heavenly equation is provided by the following 
\begin{theo}[Jones and Tod\cite{JT85}]
\label{prop_JT}
Let $({\cal M}, [g])$ be a real
four-manifold with  ASD conformal curvature,
and a  conformal non-null Killing vector.
The space of trajectories of this vector is equipped with an EW structure 
defined by
\be
\label{EWs}
h:=|K|^{-2}{g}\pm|K|^{-4}K\odot K,\qquad \om:=2|K|^{-2}
\ast_g({K}\wedge \d{K}),
\ee
where $\ast_g$ is taken w.r.t some $g\in [g]$,
$K$ is the one-form dual to the conformal Killing vector, 
and $|K|^2=g(K, K)$.
All three-dimensional EW structures arise in this way.
The $+$ and $-$ signs in {\em(\ref{EWs})} refer to the signature of $[g]$ 
being Euclidean or neutral respectively.
\end{theo}
This result was  improved in \cite{CP} and \cite{DMT00}, where it was shown
that all EW spaces can be obtained as reductions form 
scalar-flat K\"ahler, or hyper-complex four manifolds respectively.

 If we assume that there exists $g\in[g]$ such that 
$({\cal M}, g)$ is Ricci flat, so that $g$ arises form a 
solution to the heavenly equation 
(\ref{secondh}), then a connection with the 
special classes of EW spaces can be established:
\begin{itemize}
\item If the symmetry fixes all self--dual two forms 
then  the heavenly equation  reduces to the Laplace equation 
in three dimensions \cite{FP79}. The metric is in the Gibbons--Hawking class,
the resulting Einstein--Weyl structures are
trivial, and their mini-twistor space is $T\CP^1$.
\item If the symmetry rotates the self-dual two-forms, then its lift to the 
bundle of self-dual two-forms has a fixed point. If this point corresponds
to a non-simple two--form then  
the heavenly equation reduces to the BFP equation (\ref{BFP}) \cite{FP79}.
If the fixed two-form is simple, then the reduced equation is dKP (\ref{dkp}) 
\cite{DMT00}.
 \item If the symmetry is only conformal but it fixes the self-dual 
two-forms, the heavenly equation reduces to equation (\ref{PMA}) \cite{D03}.
More general conformal symmetries have been studied in \cite{DT01}.
\end{itemize}
\section{Integrability of the Hyper--Heavenly equations?}

Hyper--heavenly (HH) equations and their reduction do not enjoy the elegant
twistor description \cite{Pe76} associated to the 
anti-self-duality, and they are
believed not to be integrable. 
This may be true for  general HH spaces, 
but the simplest HH space-- the null K\"ahler 
Einstein equation (\ref{HHspace})-- shares 
an integrable root with the heavenly equation (\ref{secondh}).
To see it define $L=\Th_{xx}, M=\Th_{xy}, N=\Th_{yy}$ and write a system
of three equations resulting 
from differentiating (\ref{secondh}) w.r.t $xx, xy$ and $yy$. This system
should be complemented by adding the integrability conditions  
$L_y=M_x, M_y=N_x$ which guarantee that $L, M, N$ admit a potential
$\Theta$. The analogous procedure applied to (\ref{HHspace}) yields
the same over-determined system. The difference arises when one chooses 
the constants of integration (function of two variables) leading to back to 
$\Theta$.

There is more evidence of integrability  associated to 
hyper--heavenly spaces: In 
\cite{PrzBa84, PrzBi87} it was demonstrated that all 
Riemannian  $HH$ 
spaces of type $[D]\times$[Any] can locally be found from solutions
to the BFP equation (\ref{BFP}). Note that in this case the existence 
of the Killing vector does not have to be imposed, but it follows form the
field equations -- a product of two spinors defining a type $D$ solution
is a Killing spinor, and a contracted covariant derivative of a 
Killing spinor is a Killing vector.

In this section we shall consider the natural
one--symmetry reduction of the null--K\"ahler Einstein spaces
(\ref{HHspace}) and show that the resulting PDE in three dimensions differs 
form the dKP (\ref{dkp}) equation by a function of one variable.

Consider a symmetry $K$ which preserves the metric $g$, as well as the
nilpotent endomorphism $N$. The canonical form of such
symmetry in the coordinates adopted to the metric (\ref{nkmetric})
turns out to be $K=\p/\p w-2w\p/\p y$. This is a special 
form of Killing vector for non-expanding HH spaces, and so it must
be contained in the classification of \cite{FP78} or
\cite{Prz90}. The Killing equations yield $({\cal L}_K\Th)_{xx}=
({\cal L}_K\Th)_{yy}=0, ({\cal L}_K\Th)_{xy}=1$. 
They integrate to
\[
\label{dKPtheta}
\Th=wxy+yA(w, z)+xB(w, z)+C(w, z) +G(x, z, y+w^2).
\]
The function $C$ is pure gauge and can be set to zero without loss of
generality.
Imposing (\ref{HHspace}) and reabsorbing one arbitrary function of $z$
into $R$ (which itself can be arbitrary)  yields
\[
R+w^2-A_z-B_w=w^2\gamma(z),\qquad Q=1+\gamma(z),\qquad P=\delta(z)
\]
(where $\gamma=\gamma(z)$ and $\delta=\delta(z)$ are some arbitrary functions), and a nonlinear 
equation
\[
\label{dKPlegandre}  
-u\gamma-x\delta+G_{zu}+G_{xx}G_{uu}-G_{xu}^2=0
\qquad\mbox{where}\qquad u=y+w^2.
\]
Write this equation  as a closed system 
\begin{eqnarray}
\d G&=&G_u\d u+G_z\d z+G_x\d x,\nonumber\\
 \label{legen1}
 0&=&- (u\gamma(z)+x\delta(z))\d x\wedge\d z\wedge \d u -\d G_u\wedge\d x\wedge \d u
 -\d G_x\wedge\d G_u\wedge \d z.
 \end{eqnarray}
Now express the first equation as
$
\d(G-uG_u)=G_z\d z+G_x\d x-u\d G_u,
$
and perform a Legendre transform  
\[
p:=G_u,\qquad u=u(z, x, p),\qquad
H(z, x, p):=-G(z, x, u(z, x, p))+pu(z, x, p).
\]
The relation 
$
\d H=H_z\d z+H_x\d x+H_p\d p
$
implies $H_z=-G_z, H_x=-G_x, H_p=u$.
Equation (\ref{legen1}) yields
\[
-(H_p\gamma(z)+x\delta(z))\d x\wedge\d z\wedge\d H_p-\d p\wedge\d x\wedge\d H_p+\d H_x\wedge\d p
\wedge\d z=0,
\]
which is equivalent to 
$
(\gamma(z)H_p+\delta(z)x)H_{pp}+H_{pz}+H_{xx}=0.
$
Taking the $p$ derivative of this equation and using
$H_p=u$ gives
\[
\label{intermedia}
(-u_{z}-(\gamma(z)u+\delta(z)x)u_{p})_p=u_{xx}.
\]
If $\gamma=0$ then the above equation is linear. 
If $\gamma(z)\neq 0$ 
for some $z$ then we restrict the domain
of $z$ such that $\gamma\neq 0$ for all $z$, and define
$U(z, x, p)=u(z, x, p)+ x\delta(z)/\gamma(z)$. Finally rename the coordinates
$T=-z, X=p, Y= x$. 
To sum up, the $U(1)$--invariant null K\"ahler Einstein condition 
(\ref{HHspace}) can be reduced to a single PDE
\be
\label{genrDKP}
(U_T-\gamma(T)UU_X)_X=U_{YY}.
\ee
This can be regarded as a variable coefficient generalisation 
the dKP equation (\ref{dkp}). (Compare this reduction  
with the ASD null K\"ahler 
condition (\ref{md}) which reduces 
down to a pair of coupled integrable PDEs: the dKP and its linearisation 
\cite{D02}.) Equation (\ref{genrDKP}) admits many explicit solutions, and
shares some `integrable properties' of the dKP.
The metric on the space of orbits of the symmetry can be easily expressed in 
terms of $U$ and it derivatives, but the associated 
geometry is unclear. Its characterisation 
could shed more light on the question of
integrability of the simple $HH$ space (\ref{HHspace}).

It is just a beginning of the story, as symmetry reductions
can be performed for all hyper-heavenly spaces. This motivates
the following 
\vskip5pt
{\bf Question}
{\em What geometric structure is induced on a space of orbits of a symmetry
in a hyper-heavenly manifold?}
\section*{Acknowledgements}
Our research was partly supported by NATO grant PST.CLG.978984.


\begin{thebibliography}{jafsdl}
\frenchspacing 


\bibitem{B00} Bryant, R. L. (2000)
Pseudo-Riemannian metrics with parallel spinor fields
and vanishing Ricci tensor. Global analysis and harmonic analysis,
53--94, Semin. Congr., {\bf 4}, Soc. Math. France, Paris.

\bibitem{CP} Calderbank, D. M. J.; Pedersen, H. (2000) 
Selfdual spaces with complex structures, Einstein-Weyl geometry and geodesics.  Ann. Inst. Fourier (Grenoble)  50, no. 3, 921--963.

\bibitem{C43} Cartan, E. (1943) Sur une classe d'espaces de Weyl,
Ann. Sci. Ecole Norm. Supp. {\bf 60}, 1-16.

\bibitem{D03}  Dunajski, M. (2004) A class of Einstein--Weyl spaces associated to an integrable system of hydrodynamic type, 
J. Geom. Phys. {\bf 51}, 126-137.

\bibitem{D02}   Dunajski, M. (2002) 
Anti-self-dual four-manifolds with a parallel real spinor,  
Proc. Roy. Soc. Lond. {\bf A 458}, 1205-1222.


\bibitem{DMT00} Dunajski, M. Mason, L.J., \& Tod, K.P. (2001) 
{Einstein--Weyl geometry, the dKP equation and twistor theory},
J. Geom. Phys. {\bf 37}, 63-92.


\bibitem{DT01} Dunajski, M. \& Tod K.P. (2001) {Einstein--Weyl Structures from 
Hyper--K\"ahler Metrics with conformal Killing Vectors}, 
Diff. Geom. Appl. {\bf 14}, 39-55.

\bibitem{DT02} Dunajski, M., \& Tod, K.P. (2002) 
Einstein--Weyl spaces and  dispersionless Kadomtsev-Petviashvili 
equation from Painlev\'e I and II,
Phys. Lett {\bf 303A}, 253--264.


\bibitem{FK03} Ferapontov, E.V. \& Khusnutdinova, K.R. (2003), On the 
integrability of $2+1$-dimensional quasilinear systems,
nlin.SI/0305044; to appear in Comm. Math. Phys.


\bibitem{FP78} Finley, J. D. \& Pleba\'nski, J. F. (1978) Killing vectors in plane HH spaces.  J. Mathematical Phys.  19, no. 4, 760--766.

\bibitem{FP79} Finley, J.D.  \& Pleba\'nski, J.F. (1979) 
The classification of all ${\cal H}$ spaces admitting a Killing vector, 
J. Math. Phys. {\bf 20}, 1938.

\bibitem{GMM} Guil, F., Manas, M., \and  Martinez Alonso, L. (2003)
The Whitham hierarchies: reductions and hodograph solutions.  
J. Phys. A 36,  no. 14, 4047--4062.

\bibitem{Hi82} Hitchin, N. (1982) Complex manifolds and Einstein's
equations, in {\em Twistor Geometry and Non-Linear systems}, LNM 970,
ed. Doebner, H.D. \& Palev, T.D.
Adv. in Math. {\bf 97} 74-109. 


\bibitem{JT85}Jones, P. \&  Tod, K.P. (1985) Minitwistor spaces and
Einstein-Weyl spaces, Class. Quantum Grav. {\bf 2} 565-577.

\bibitem{Ko63} Kodaira, K. (1963)
On stability of compact submanifolds of complex manifolds,
 Am. J. Math.  {\bf 85}, 79-94.

\bibitem{L91} LeBrun, C.R. (1991)
Explicit self-dual metrics on 
$\CP^2 \# \cdots \# \CP^2$, J. Diff. Geom. {\bf 34}  233-253.

\bibitem{Pe76} Penrose, R. (1976) Nonlinear gravitons and curved
twistor theory, Gen. Rel. Grav. {\bf 7}, 31-52.
 


\bibitem{Pl75} Pleba\'nski, J. F. (1975) Some solutions of complex
Einstein Equations, J. Math. Phys. {\bf 16} 2395-2402.

\bibitem{PR76}
Pleba\'nski, J. F.\& Robinson, I. (1976) 
Left-degenerate vacuum metrics.  
Phys. Rev. Lett.  37  (1976), no. 9, 493--495. 

\bibitem{PP89} Pleba\'nski, J. F. \& Przanowski, M. (1989)
Two-sided conformally recurrent four-dimensional Riemannian manifolds.  J. Math. Phys.  30,  no. 9, 2101--2109

\bibitem{Prz90} Przanowski, M. (1990) 
The group theoretic analysis of hyperheavenly equations.  
J. Math. Phys.  31,  no. 3, 653--658.

\bibitem{PrzBa84} Przanowski, M.\& Baka, B. (1984)
One-sided type-$D$ gravitational instantons.  Gen. Relativity Gravitation  16  (1984),  no. 9, 797--803. 

\bibitem{PrzBi87}
Przanowski, M. \& Bialecki, S. 
(1987)  Lie--B\"acklund Transformation and Gravitational Instanton, Acta Phys. Polon. {\bf B} 18 no. 10, 879--899.  


\bibitem{T95} Tod, K. P. (1995) Scalar-flat K{\"a}hler Metrics from
Painlev{\'e}-III,
Class. Quantum Grav. {\bf 12} 1535-1547




\end{thebibliography}
\end{document}